\documentclass[aps, preprint, groupedaddress, floatfix, longbibliography]{revtex4-2}

\usepackage{epsfig}
\usepackage{epstopdf}
\usepackage{graphicx}
\usepackage{amsmath}
\usepackage{color}
\usepackage[export]{adjustbox}
\usepackage[colorlinks=true, allcolors=blue]{hyperref}
\usepackage{lineno}
\begin{document}

\title{Three-Dimensional Fermi Surface, Van Hove Singularity and Enhancement of Superconductivity in Infinite-Layer Nickelates}

\author{Chengliang Xia$^1$, Shengjie Zhou$^1$ and Hanghui Chen$^{1,2}$\footnote{hanghui.chen@nyu.edu}}
\affiliation{$^1$NYU-ECNU Institute of Physics, NYU Shanghai, Shanghai 200122, China\\
  $^2$Department of Physics, New York University, New York, New York 10012, USA}

\date{\today}

\begin{abstract}
Recent experiments reveal a three-dimensional (3D) Fermi surface with a clear $k_z$ dispersion in infinite-layer nickelates, distinguishing them from their cuprate superconductor counterparts. However, the impact of this difference on the superconducting properties of nickelates remains unclear. Here, we employ a combined random-phase-approximation and dynamical-mean-field-theory (RPA+DMFT) approach to solve the linearized gap equation for superconductivity. We find that, compared to the cuprate-like two-dimensional (2D) single-orbital Fermi surface, the van Hove singularities on the 3D Fermi surface of infinite-layer nickelates strengthen spin fluctuations by driving the system closer to antiferromagnetic instabilities, thereby significantly enhancing superconductivity. Our findings underscore the critical role of the van Hove singularities in shaping the superconducting properties of infinite-layer nickelates and, more broadly, highlight the importance of subtle Fermi surface features in modeling material-specific unconventional superconductors.
\end{abstract}

\maketitle


\newpage

The discovery of superconductivity in infinite-layer nickelates has sparked significant interest in the condensed matter community~\cite{Li2019,PhysRevLett.125.147003,PhysRevLett.125.027001,PhysRevX.11.011050,Wang2021,PhysRevLett.126.197001,Lu2021,PhysRevB.102.020502,PhysRevB.102.100501,PhysRevB.101.041104,PhysRevB.105.115134,PhysRevLett.126.127401,PhysRevResearch.1.032046,Rossi2022,Tam2022,Ding2023,10.1093/nsr/nwae194,Liu2020a,Gu2020a,PhysRevLett.124.207004,PhysRevB.102.241112,PhysRevB.102.220501,PhysRevB.100.201106,PhysRevB.102.161118,PhysRevB.103.045103,10.1093/nsr/nwaa218,doi:10.1073/pnas.2007683118,PhysRevResearch.2.013214,PhysRevResearch.2.023112,PhysRevLett.133.066503,PhysRevLett.126.087001,PhysRevB.109.184505,PhysRevLett.129.027002,PhysRevB.106.134504,PhysRevB.111.045151}. While similarities between infinite-layer nickelates and cuprate superconductors were recognized early on~\cite{PhysRevX.10.011024,PhysRevX.10.021061}, increasing attention has been drawn to their differences~\cite{PhysRevB.102.195117,10.3389/fphy.2022.835942,PhysRevX.10.041047,PhysRevX.10.041002,PhysRevLett.129.077002}. Notably, infinite-layer nickelates possess a conduction band that crosses the Fermi level~\cite{PhysRevB.70.165109,Hepting2020,Gu2020}, giving rise to the so-called ``self-doping'' effect~\cite{PhysRevB.101.020501,PhysRevB.101.081110}. Recent angle-resolved photoemission spectroscopy (ARPES) measurements have revealed the Fermi surface of superconducting La$_{0.8}$Sr$_{0.2}$NiO$_2$, consisting of a large sheet derived from the Ni-$d_{x^2-y^2}$ orbital and an electron pocket at the Brillouin zone corner~\cite{doi:10.1126/sciadv.adr5116}. A key distinction between infinite-layer nickelates and cuprate superconductors is that in the former, the Ni-$d_{x^2-y^2}$ derived Fermi surface is three-dimensional with a clear $k_z$ dispersion~\cite{doi:10.1126/sciadv.adr5116,PhysRevB.101.060504,PhysRevB.101.241108,PhysRevResearch.6.043104}, while the latter has a two-dimensional Cu-$d_{x^2-y^2}$ derived Fermi surface~\cite{RevModPhys.75.473,Keimer2015}. A fundamental open question is whether these subtle features in the low-energy electronic structure may impact the superconducting properties of infinite-layer nickelates.

In this work, we directly address this question by comparing two widely used low-energy effective models for infinite-layer nickelates. The first is a two-orbital $ds$ model~\cite{10.3389/fphy.2022.835942,Chen2023,doi:10.1126/sciadv.adr5116,PhysRevB.108.245115,PhysRevB.106.224517} which includes a correlated Ni-$d_{x^2-y^2}$ orbital with an onsite Hubbard interaction and an effective $s$ orbital. Crucially, the hybridization between these orbitals is essential to capture the three-dimensional character of the Fermi surface~\cite{Chen2023,doi:10.1126/sciadv.adr5116}. The second is the single-orbital two-dimensional Hubbard model~\cite{Kitatani2020,10.3389/fphy.2022.834682,PhysRevResearch.6.043104}, referred to in this work as the ``$d$-only'' model. To study superconductivity, we combine the random-phase approximation (RPA)~\cite{PhysRevB.31.4403,Graser_2009,PhysRevB.79.224511,RevModPhys.84.1383} with dynamical mean-field theory (DMFT)~\cite{RevModPhys.68.13,RevModPhys.78.865} to construct the pairing potential and solve the linearized gap equation. This approach allows us to access the full range of interaction strengths, from weak to strong coupling~\cite{PhysRevB.95.174504,PhysRevLett.129.077002,PhysRevLett.123.247001}. We find that, for the same $d$-orbital occupancy and interaction strength, the $ds$ model yields a significantly larger superconducting eigenvalue than the $d$-only model, suggesting a higher superconducting transition temperature. This difference becomes more pronounced at intermediate to strong interaction strengths, which are most relevant for infinite-layer nickelates~\cite{PhysRevLett.125.077003,PhysRevB.100.205138,Kitatani2020,doi:10.1126/sciadv.adr5116}. The enhancement arises from van Hove singularities (VHS) on the three-dimensional Fermi surface~\cite{PhysRevB.107.075159}, which are absent in the simplified $d$-only model. These subtle features drive the system closer to a spin-density-wave (SDW) instability, substantially enhancing spin fluctuations and, consequently, superconductivity. Our results underscore the importance of subtle Fermi surface features in modeling material-specific unconventional superconductors.     

We perform density functional theory (DFT) calculations~\cite{PhysRev.136.B864,PhysRev.140.A1133} on the prototypical infinite-layer nickelate La$_{0.8}$Sr$_{0.2}$NiO$_2$, and downfold the electronic structure to two low-energy effective models: the $ds$ model and the $d$-only model. In both models, we apply an onsite Hubbard interaction to the Ni-$d_{x^2-y^2}$ orbital and use DMFT to extract the quasi-particle dispersion $E(\textbf{k})$ and quasi-particle weight $Z_d$ for the Ni-$d_{x^2-y^2}$ orbital. Based on these quantities, we construct the pairing potential and solve the linearized gap equation using a combined RPA and DMFT (RPA+DMFT) method~\cite{PhysRevB.95.174504,PhysRevLett.123.247001,PhysRevLett.129.077002}. Finally, we compute the leading superconducting eigenvalue and compare the results between the two models. Further methodological details are provided in the Supplementary Notes 1-8.

\begin{figure}[t]
\includegraphics[angle=0,width=0.9\textwidth]{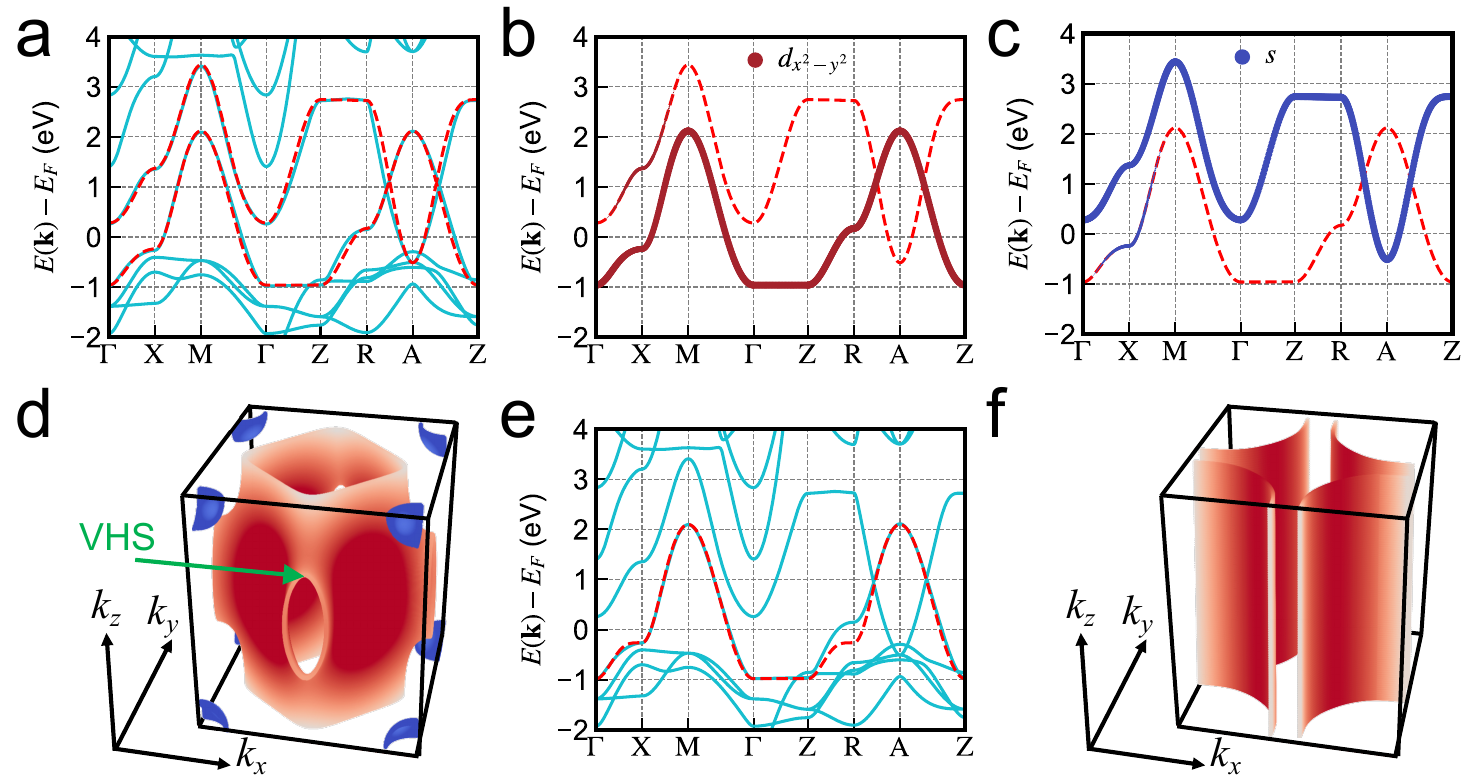}
\caption{\label{fig1} (a) Comparison between the DFT band structure of La$_{0.8}$Sr$_{0.2}$NiO$_2$ (cyan solid line) and the band structure of the tight-binding part of the $ds$ model (red dashed line). (b, c) Band structure of the tight-binding part of the $ds$ model (red dashed lines). (b) Dark red dots indicate the projection onto the Ni-$d_{x^2-y^2}$ orbital. (c) Blue dots indicate the projection onto the effective $s$ orbital. (d) Fermi surface of the $ds$ model, with color indicating orbital character: dark red for the Ni-$d_{x^2-y^2}$ and blue for the effective $s$ orbital. A representative van Hove singularity (VHS) is marked by a green arrow. (e) Comparison between the DFT band structure (cyan solid line) and the tight-binding band structure of the $d$-only model (red dashed line). (f) Fermi surface of the $d$-only model. The dark red shading indicates projection onto the Ni-$d_{x^2-y^2}$ orbital.
}
\end{figure}

Figure~\ref{fig1}(a) presents the electronic structure of La$_{0.8}$Sr$_{0.2}$NiO$_2$. The cyan solid lines represent the DFT-calculated band structure, while the red dashed lines show the band structure reproduced by the tight-binding part of the $ds$ model. Fig.~\ref{fig1}(b) and \ref{fig1}(c) display the orbital projections onto the Ni-$d_{x^2-y^2}$ and the effective $s$ orbitals, respectively. The band originating from the Ni-$d_{x^2-y^2}$ orbital (dark red) crosses the Fermi level throughout the entire Brillouin zone, whereas the conduction band (blue) intersects the Fermi level only near the zone corner $A$ point $(\pi, \pi, \pi)$.  Fig.~\ref{fig1}(d) shows the corresponding Fermi surface obtained from the $ds$ model. The red surface represents the large Fermi sheet derived from the Ni-$d_{x^2-y^2}$ orbital, and the blue pocket corresponds to the conduction band near the $A$ point. Our calculated Fermi surface aligns well with the recent ARPES measurements~\cite{doi:10.1126/sciadv.adr5116}. Due to hybridization between the Ni-$d_{x^2-y^2}$ and the effective $s$ orbitals, the $ds$ model captures a key feature: the Ni-$d_{x^2-y^2}$-derived Fermi surface appears hole-like in the $k_z = 0$ plane but becomes electron-like in the $k_z = \pi$ plane. This change in the $k_z$ dispersion inevitably gives rise to van Hove singularities (VHS) on certain $k_z$ slices, one of which is highlighted by the green arrow in Fig.~\ref{fig1}(d). For comparison, we also consider a $d$-only model. The red dashed lines in Fig.~\ref{fig1}(e) show the corresponding tight-binding band structure, and the resulting Fermi surface is displayed in Fig.~\ref{fig1}(f). In contrast to the $ds$ model, the $d$-only model yields a two-dimensional Fermi surface with no $k_z$ dispersion, matching the $ds$ model's Fermi surface at $k_z = 0$. Notably, VHS are absent on the Fermi surface of the $d$-only model.

\begin{figure}[t]
\includegraphics[angle=0,width=0.9\textwidth]{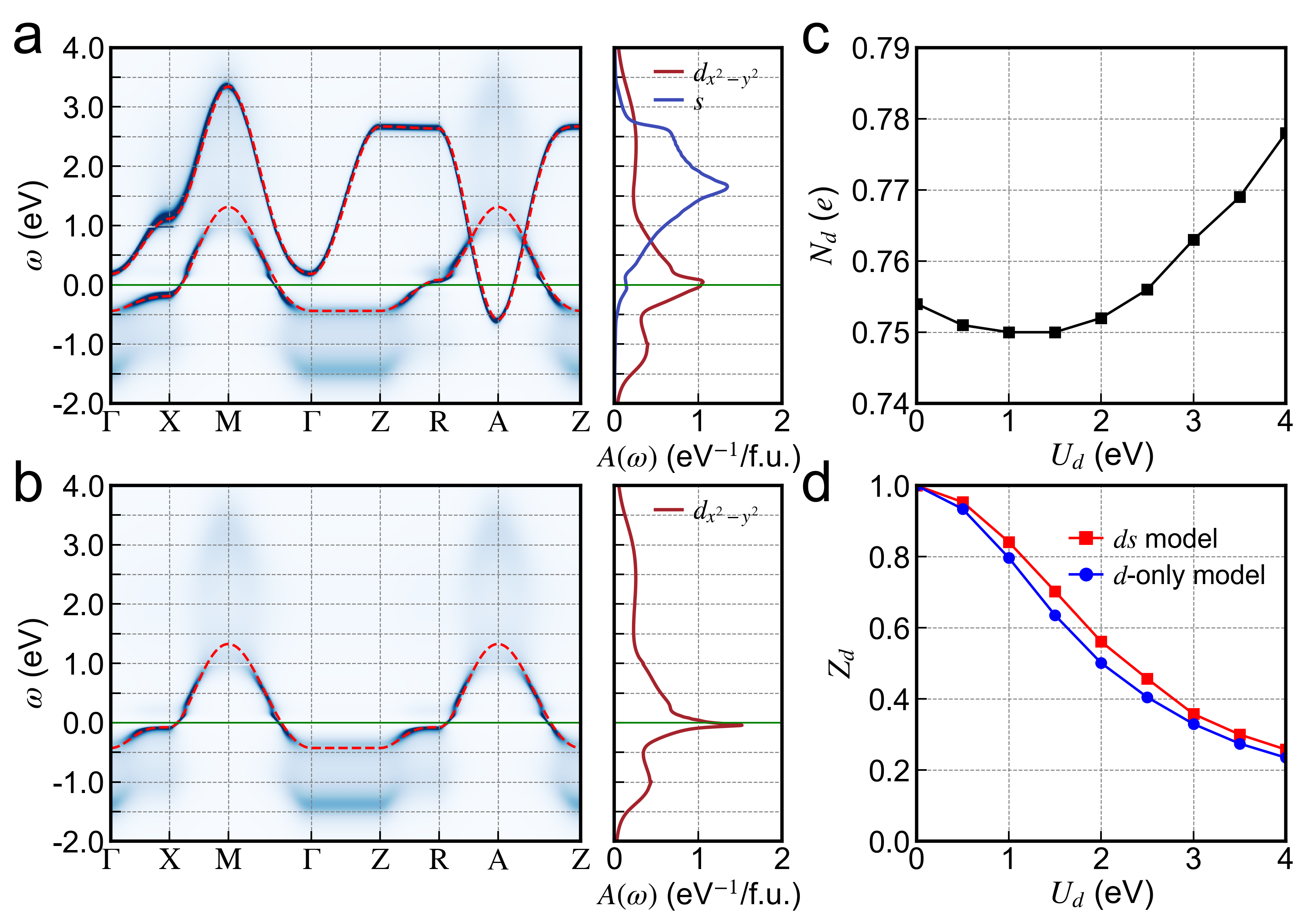}
\caption{\label{fig2} 
(a) Left: DMFT momentum-resolved spectral function $A(\mathbf{k},\omega)$ of the $ds$ model calculated at $U_d = 2.5$ eV. The red dashed lines indicate the quasi-particle energy. Right: DMFT orbital-resolved spectral function $A_{\nu}(\omega)$ of the $ds$ model. The dark red and blue lines represent the projections onto the Ni-$d_{x^2 - y^2}$ and effective $s$ orbitals, respectively. (b) Left: DMFT momentum-resolved spectral function $A(\mathbf{k},\omega)$ of the $d$-only model at $U_d = 2.5$ eV. The red dashed line indicates the quasi-particle energy. Right: DMFT spectral function $A(\omega)$ of the $d$-only model. The dark red line corresponds to the projection onto the Ni-$d_{x^2 - y^2}$ orbital. (c) Occupancy of the Ni-$d_{x^2 - y^2}$ orbital in the $ds$ model as a function of $U_d$. (d) Quasi-particle weight $Z_d$ of the Ni-$d_{x^2 - y^2}$ orbital as a function of $U_d$, shown for both the $ds$ model (red symbols) and the $d$-only model (blue symbols).
}
\end{figure}

Next we employ the DMFT method to calculate the many-body electronic structure of the two models and extract the quasi-particle energy $E(\textbf{k})$. The left panel of Figure~\ref{fig2}(a) is the momentum-resolved spectral function $A(\textbf{k},\omega)$ of the $ds$ model calculated at $U_d$ = 2.5 eV. The peak position of $A(\textbf{k},\omega)$ corresponds to the quasi-particle energy. Then we renormalize the onsite energies and hopping parameters in the tight-binding part of the $ds$ model to fit the quasi-particle energy. The resulting electronic structure is shown by the red dashed lines. The subsequent superconductivity calculations will be based on the tight-binding part of the ``dressed'' $ds$ model that fits the DMFT-calculated quasi-particle energy (as opposed to the ``bare'' $ds$ model that fits the DFT-calculated band structure). The right panel of Fig.~\ref{fig2}(a) is the orbital-resolved spectral function $A_{\nu}(\omega)$. The dark red and blue curves correspond to projections onto the Ni-$d_{x^2-y^2}$ and the effective $s$ orbitals, respectively. The spectral function of the Ni-$d_{x^2-y^2}$ orbital exhibits a pronounced quasi-particle peak at the Fermi level. The spectral function of the effective $s$ orbital reveals a weak occupation due to an electron pocket at $A$ point, indicating the presence of the ``self-doping'' effect. Fig.~\ref{fig2}(b) shows analogous results for the $d$-only model: the momentum-resolved spectral function $A(\mathbf{k}, \omega)$ (left panel) and the orbital-resolved spectral function $A_{\nu}(\omega)$ (right panel), both computed at $U_d = 2.5$~eV and $N_d = 0.756$ (explained below). The red dashed lines represent the quasi-particle band structure of the ``dressed'' $d$-only model, constructed by renormalizing its tight-binding parameters to match the DMFT-derived quasi-particle energies. A sharp quasi-particle peak is observed at the Fermi level. Fig.~\ref{fig2}(c) displays the $d$-orbital occupancy $N_d$ in the $ds$ model. Since Ni-$d_{x^2-y^2}$ orbital is nominally half-filled in infinite-layer nickelates, the physical total occupancy is 1, which is partitioned into the Ni-$d_{x^2-y^2}$ orbital and the effective $s$ orbital. We find that as $U_d$ increases, $N_d$ initially decreases slightly before increasing again, indicating a suppression of the self-doping effect at stronger interactions. This trend is consistent with previous studies~\cite{Gu2020,PhysRevB.101.081110}. Since electron correlations are sensitive to $d$-orbital filling, we fix $N_d$ to the same value when comparing the two models at a given $U_d$. In Fig.~\ref{fig2}(d), we show the quasi-particle weight $Z_d$ of the Ni-$d_{x^2-y^2}$ orbital for both models as a function of $U_d$. As expected, the quasi-particle weight $Z_d$ decreases monotonically with increasing $U_d$, indicating proximity to a Mott transition~\cite{Gu2020,PhysRevB.101.081110,PhysRevB.90.195114}. Notably, at the same $U_d$, the $ds$ model exhibits a slightly larger $Z_d$ than the $d$-only model, reflecting the enhanced screening due to the additional $s$ orbital in the $ds$ model.


\begin{figure}[t]
\includegraphics[angle=0,width=0.9\textwidth]{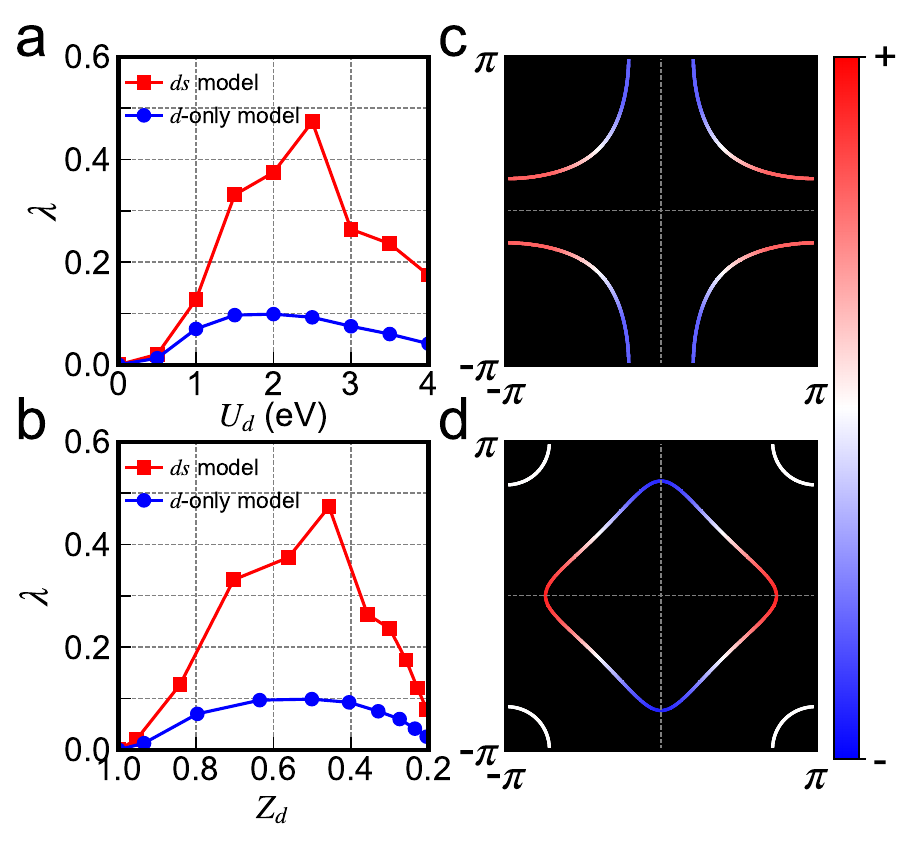}
\caption{\label{fig3} (a) Leading superconducting eigenvalue of the linearized gap equation as a function of $U_d$. The pairing potential is based on the RPA+DMFT method. The red and blue symbols correspond to the $ds$ model and the $d$-only model, respectively. (b) Same as (a), but plotted as a function of the quasi-particle weight $Z_d$ of the Ni-$d_{x^2-y^2}$ orbital. (c, d) Gap function of the leading superconducting instability in the $ds$ model at $U_d$ = 2.5 eV, exhibiting $d_{x^2-y^2}$ symmetry. (c) Gap function in the $k_z$ = 0 plane. (d) Gap function in the $k_z = \pi$ plane.}
\end{figure}

Now equipped with the quasi-particle energy $E(\textbf{k})$ and quasi-particle weight $Z_d$, we proceed to solve the linearized gap equation using the RPA+DMFT method~\cite{PhysRevB.95.174504,PhysRevLett.123.247001,PhysRevLett.129.077002}. This approach allows us to explore a broad range of interaction strengths, from weak to strong coupling, extending the standard RPA method, which is limited to the weak-coupling regime~\cite{Graser_2009,PhysRevB.92.104505}. Figure~\ref{fig3}(a) shows the leading superconducting eigenvalue obtained from the linearized gap equation for both models. We find that the eigenvalue of the $ds$ model is significantly larger than that of the $d$-only model across the range $U_d = 1.5 - 4.0$ eV. Given that the nearest-neighbor hopping of Ni-$d_{x^2-y^2}$ orbital is $t_d = 0.39$ eV, this corresponds to an intermediate-to-strong interaction strength ($U_d/t_d \sim 4-10$), which is relevant to infinite-layer nickelates~\cite{doi:10.1126/sciadv.adr5116,PhysRevLett.125.077003,PhysRevB.100.205138,Kitatani2020}. Since at the same $U_d$ value, the quasi-particle weight $Z_d$ of the Ni-$d_{x^2-y^2}$ orbital differs slightly between the two models (as shown in Fig.~\ref{fig2}(d)), a more physically meaningful comparison is to examine the superconducting eigenvalue as a function of $Z_d$. This result is shown in Fig.~\ref{fig3}(b). Consistent with Fig.~\ref{fig3}(a), the $ds$ model exhibits a substantially larger eigenvalue than the $d$-only model over a wide range of quasi-particle weights, including the experimentally observed values $Z_d \sim 0.3 - 0.5$~\cite{doi:10.1126/sciadv.adr5116}. For all investigated values of $U_d$, we find that the most favorable pairing symmetry in both models is $d_{x^2-y^2}$. The corresponding superconducting gap function of the $ds$ model in the $k_z = 0$ and $k_z = \pi$ planes is displayed in Fig.~\ref{fig3}(c) and (d), respectively. For completeness, we also compare the two models using the standard RPA method (see Supplementary Note 9) and find that the leading superconducting eigenvalue of the $ds$ model is always larger than that of the $d$-only model, when $U_d$ is below the critical value for SDW. This is qualitatively consistent with the RPA+DMFT results.

\begin{figure}[t]
\includegraphics[angle=0,width=0.9\textwidth]{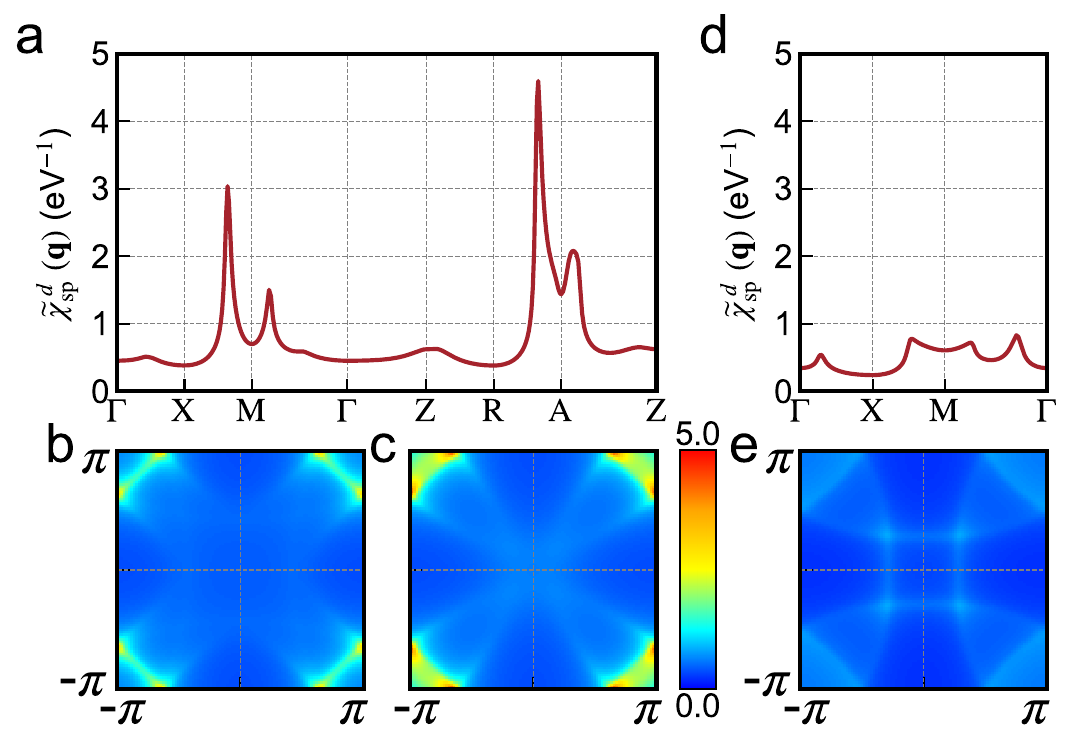}
\caption{\label{fig4} (a) Spin susceptibility $\widetilde{\chi}^d_{\rm{sp}}(\mathbf{q})$ of the Ni-$d_{x^2 - y^2}$ orbital in the $ds$ model along the high-symmetry \textbf{q}-path, calculated using the RPA+DMFT method at $U_d = 2.5$ eV. (b, c) Spin susceptibility $\widetilde{\chi}^d_{\rm{sp}}(\mathbf{q})$ of the Ni-$d_{x^2 - y^2}$ orbital in the $ds$ model in the $q_z = 0$ plane (b) and $q_z = \pi$ plane (c). (d) Spin susceptibility $\widetilde{\chi}^d_{\rm{sp}}(\mathbf{q})$ of the $d$-only model along the high-symmetry \textbf{q}-path, calculated using the RPA+DMFT method at $U_d = 2.5$ eV. (e) Corresponding spin susceptibility $\widetilde{\chi}^d_{\rm{sp}}(\mathbf{q})$ of the $d$-only model in the $q_z = 0$ plane.
}
\end{figure}


Finally, we clarify why the $ds$ model exhibits a superconducting eigenvalue substantially larger than that of the $d$-only model. As previously discussed, the $ds$ model features a three-dimensional Fermi surface, where the Ni-$d_{x^2-y^2}$-derived band is hole-like in the $k_z = 0$ plane and electron-like in the $k_z = \pi$ plane. This topology enforces the presence of VHS on the Fermi surface at an intermediate $k_z$ plane. The existence of VHS on the Fermi surface drives the system closer to SDW instabilities, thereby significantly amplifying spin fluctuations. To illustrate this explicitly, Figure~\ref{fig4}(a) shows the momentum-resolved spin susceptibility $\widetilde{\chi}^{d}_{\rm{sp}}(\mathbf{q})$ of the Ni-$d_{x^2-y^2}$ orbital in the $ds$ model, calculated at $U_d = 2.5$ eV along a high-symmetry $\mathbf{q}$-path. The corresponding distributions in the $q_z = 0$ and $q_z = \pi$ planes are shown in Fig.~\ref{fig4}(b) and (c), respectively. Prominent peaks around $M(\pi, \pi, 0)$ and $A(\pi, \pi, \pi)$ points indicate strong antiferromagnetic spin fluctuations. In contrast, the $d$-only model possesses a purely two-dimensional Fermi surface and lacks the VHS feature. Fig.~\ref{fig4}(d) and (e) display $\widetilde{\chi}^{d}_{\mathrm{sp}}(\mathbf{q})$ for the $d$-only model at the same $U_d$ and $N_d$. The spin susceptibility around the $M$ point is noticeably smaller and shows no sharp peaks, underscoring substantially weaker magnetic fluctuations compared to the $ds$ model. The full evolution of $\widetilde{\chi}^{d}_{\mathrm{sp}}(\mathbf{q})$ across the entire range of $U_d$ for both models is provided in Supplementary Note 10. In addition to the direct analysis of spin susceptibility $\widetilde{\chi}^{d}_{\mathrm{sp}}(\mathbf{q})$, an alternative way to quantify spin fluctuations is to evaluate the system's proximity to SDW instabilities by calculating the product of the susceptibility and the interaction strength, $\widetilde{\chi}^{d}_0(\mathbf{q}) U_d$~\cite{Graser_2009,lechermann2023electronic}. We find that, for $U_d$ in the range of 1.5--4.0~eV, the maximum value of $\widetilde{\chi}^{d}_0(\mathbf{q}) U_d$ in the $ds$ model is substantially closer to unity than in the $d$-only model, indicating a stronger tendency toward magnetic ordering and enhanced spin fluctuations. The full evolution of $\widetilde{\chi}^{d}_0(\mathbf{q}) U_d$ as a function of $U_d$ for both models is also presented in Supplementary Note~10.


In conclusion, by solving the linearized gap equation using the RPA+DMFT method, we demonstrate that the three-dimensional Fermi surface of infinite-layer nickelates substantially enhances superconductivity compared to the simplified two-dimensional single-orbital Fermi surface. This enhancement originates from the presence of van Hove singularities on the three-dimensional Fermi surface, which brings the system closer to spin-density-wave instabilities and markedly amplifies antiferromagnetic spin fluctuations. Our results highlight that subtle features of the Fermi surface play a pivotal role in shaping the superconducting properties of infinite-layer nickelates. More broadly, our work underscores the importance of incorporating sufficient material-specific Fermi surface details in modeling unconventional superconductors.

\begin{acknowledgments}
  We are grateful to Peter Hirschfeld, Andreas Kreisel, Ilya Eremin and Karsten Held for useful discussions. This project was financially supported by the National Natural Science Foundation of China under project number 12374064 and 12434002, Science and Technology Commission of Shanghai Municipality under grant number 23ZR1445400 and a grant from the New York University Research Catalyst Prize under project number RB627. C.X. was supported by the National Natural Science Foundation of China under project number 12404082. NYU High-Performance-Computing (HPC) provides computational resources.
\end{acknowledgments}

\bibliographystyle{apsrev}
\bibliography{main}

\end{document}